\documentclass[twocolumn,showpacs,nofootinbib,10pt]{revtex4-1}

\usepackage{color}
\usepackage{graphicx}
\usepackage{graphicx,float}\usepackage[all]{xy}
\usepackage{amsmath,upgreek}
 \newcommand{\bltx}{\textcolor{black}}
\usepackage{amssymb}

\usepackage{textgreek}

\newcommand{\beq}{\begin{eqnarray}}
\newcommand{\eeq}{\end{eqnarray}}
\newcommand{\be}{\begin{equation}}
\newcommand{\ee}{\end{equation}}

\newcommand{\bea}{\begin{eqnarray}}
\newcommand{\eea}{\end{eqnarray}}

\newcommand{\ba}{\begin{eqnarray}}
\newcommand{\ea}{\end{eqnarray}}
\usepackage[colorlinks,hyperindex,unicode]{hyperref}
\definecolor{green1}{RGB}{0,128,0} 
\hypersetup{hidelinks,backref=true,pagebackref=true,hyperindex=true,colorlinks=true,breaklinks=true,urlcolor= blue}
\hypersetup{%
  colorlinks = true,
  linkcolor  = blue,
  citecolor = green1,
}
\usepackage{bookmark,textgreek}
\usepackage{hyperref,color,xcolor}
\hypersetup{hidelinks,hyperindex=true,colorlinks=true,breaklinks=true,urlcolor= blue}
\hypersetup{%
  colorlinks = true,
  linkcolor  = blue
}
\begin{document}
\title{The minimal of geometric deformation of Yang-Mills-Dirac  stellar configurations}

\author{Rold\~ao da Rocha}
\affiliation{Center of Mathematics,  Federal University of ABC, 09210-580, Santo Andr\'e, Brazil.}
\email{roldao.rocha@ufabc.edu.br}

%
%


\begin{abstract} 
The method of minimal geometric deformation (MGD) is used to derive static, strongly gravitating,  spherically symmetric, compact stellar distributions that are solutions of the Yang-Mills-Einstein-Dirac coupled field equations, on fluid membranes with finite tension. Their solutions  characterize MGD Yang-Mills-Dirac stars, whose mass has order of the Chandrasekhar mass, once the range of both the fermionic self-interaction and the Yang-Mills coupling constants is suitably chosen. Physical features of MGD Yang-Mills-Dirac stars are then discussed, and their ADM mass are derived, as a function of the fermion coupling constant, the finite brane tension, and the Yang-Mills running parameter as well. 

\end{abstract}


\keywords{Minimal geometric deformation; fluid branes; brane tension; stellar configurations; spinor fields; Yang-Mills fields}

\maketitle

\section{Introduction} 

The minimal geometric deformation, MGD, and the MGD-decoupling methods consist of  well-succeeded procedures that can engender analytical solutions of the brane Einstein's effective field equations, in AdS/CFT and its membrane paradigm  \cite{Ovalle:2017fgl,ovalle2007,Ovalle:2010zc,Casadio:2012rf,darkstars}. Our universe, the codimension-1 brane with intrinsic tension, is embedded in a bulk   \cite{Antoniadis:1998ig,Antoniadis:1990ew}. MGD procedures can be implemented when one deforms already known solutions in general relativity (GR). Standard gravity in GR consists of 
the rigid brane, with infinite tension, limit. The brane tension is interpreted as the finite vacuum energy. Since recent phenomenological data imply a finite brane tension ($\sigma$), with the most precise  observational bound $\sigma \gtrapprox  2.81\times10^{-6} \;{\rm GeV^4}$ \cite{Fernandes-Silva:2019fez}, 
then the MGD method represents a realistic procedure to derive and analyze compact stellar distributions, complying with current cosmological observations. The finite brane tension drives the ways to deform the Schwarzschild solution of the Einstein's field equations \cite{Casadio:2013uma,Ovalle:2013vna,Ovalle:2019qyi,Ovalle:2013xla,Ovalle:2018vmg}. 
 The CMB anisotropic aspects, showed  by WMAP, forced the fluid brane tension to satisfy the E\"otv\"os law, asserting that the surface tension of a membrane is proportional to its temperature  \cite{Abdalla:2009pg,daRocha:2012pt,Gergely:2008jr}. Regarding models describing cosmic inflation, the tension of the brane must have fluctuated, as the temperature of the universe decreased down to the current value $\sim 2.7$ K \cite{Gergely:2008jr}.

The MGD provides analytical solutions of strongly gravitating, compact star distributions and black holes as well \cite{Casadio:2012rf,darkstars,Ovalle:2007bn,ovalle2007,Ovalle:2018umz}. 
 The MGD is ruled by a running parameter, proportional to the inverse of the brane tension, with recent observational/experimental constraints \cite{Casadio:2015jva,Casadio:2016aum,Fernandes-Silva:2019fez}. The MGD and its extensions have been employed to study the configurational stability of stellar distributions \cite{Casadio:2016aum,Fernandes-Silva:2019fez}, whose observational signature in LIGO and eLISA was proposed in Ref. \cite{daRocha:2017cxu,Fernandes-Silva:2018abr}. 
 Einstein-Klein-Gordon configurations, using the MGD gravitational
                        decoupling, were studied in Ref. \cite{Ovalle:2018ans}.  The MGD procedure and the MGD gravitational decoupling method as well have been also investigated and used in Refs. \cite{Contreras:2018gzd,Casadio:2019usg,Rincon:2019jal,Ovalle:2019lbs,Gabbanelli:2019txr,Maurya:2019noq,Tello-Ortiz:2019gcl,Morales:2018urp,Morales:2018nmq,Panotopoulos:2018law,Singh:2019ktp,Maurya:2019wsk}, encompassing also the analysis and scrutiny of anisotropic configurations  \cite{Ovalle:2017wqi,Gabbanelli:2018bhs,PerezGraterol:2018eut,Heras:2018cpz,Torres:2019mee,Hensh:2019rtb,Contreras:2019iwm,Contreras:2018nfg,Sharif:2018toc,Stelea:2018cgm,Maurya:2019hds,Cedeno:2019qkf,Deb:2018ccw}. Higher derivative terms in the  gravitational decoupling were studied in Refs. \cite{Estrada:2019aeh,Sharif:2018tiz}. Besides,  the $(2+1)$-dimensional version of the MGD gravitational decoupling, with several applications,  were discussed in Refs. \cite{Contreras:2018vph,Contreras:2019fbk,Contreras:2019mhf,Sharif:2019mzv}. Still, new nuances of the MGD in AdS/CFT have been 
                        paved by the study of the holographic entanglement entropy of MGD solutions \cite{daRocha:2019pla}.

In this work, MGD Yang-Mills-Dirac  stellar configurations will be derived as solutions of Yang-Mills-Einstein-Dirac coupled system of field equations on a fluid brane with finite tension. We will analyze the finite brane tension influence on the compact stellar configurations. Without taking into account Yang-Mills fields, solutions of the Einstein-Dirac coupled system consist of Dirac stellar distributions, driven  by non-linear spinor fields~\cite{Herdeiro:2017fhv,Dzhunushaliev:2018jhj}. 
After, Maxwell, Yang-Mills and even Proca fields were coupled to the Einstein-Dirac coupled system, always 
in the GR, $\sigma\to\infty$, limit  \cite{Dzhunushaliev:2019uft,Dzhunushaliev:2019kiy,Herdeiro:2019mbz}. 
MGD Dirac stars were proposed and studied in Ref. \cite{daRocha:2020rda}, with astonishing physical consequences that arise from the brane tension finite value. 
The Einstein-Dirac coupled system of field equations 
have static and analytical solutions \cite{Bronnikov:2019nqa,Saha:2015ata}. Self-gravitating 
systems of spin-1/2 fermionic fields were also studied in Ref. \cite{Fabbri:2014zya,Fabbri:2014wda}.  The main aim of this work is to study MGD compact stellar configurations, as solutions of the effective Yang-Mills-Einstein-Dirac coupled system of field equations on a fluid brane with finite tension. 
 
Among the possible Yang-Mills fields that can be employed, we will approach 
 SU(2) monopole-like fields. In fact, Refs. \cite{Dzhunushaliev:2019uft,Dzhunushaliev:2019kiy,Herdeiro:2019mbz,Bartnik:1988am} already studied static, spherically symmetric, configurations arising from Yang-Mills equations, however only in the GR limit of an infinitely rigid brane, with $\sigma\to\infty$. Hence, it is natural to argue, in a realistic model that complies with recent observational data, how the finite tension of the brane  can drive new MGD solutions of the Yang-Mills-Einstein-Dirac system. These solutions will be shown to form compact MGD Yang-Mills-Dirac stellar configurations, whose physical properties will be scrutinized. 
Besides the MGD procedure, a MGD-decoupling-like technique will be also utilized 
to construct a more reliable stress-energy brane tensor, induced by non-linear spinor fields. 
For fermion masses that are orders of magnitude smaller than the Planck mass, the MGD-decoupled solutions  will be shown to characterize MGD Yang-Mills-Dirac stars, with mass of the order of the Chandrasekhar mass. Self-interacting spinor fields will be then employed to investigate the Yang-Mills-Dirac stars main physical features, whose ADM mass is a function of the fermion  self-interaction and Yang-Mills coupling constants.
 
This paper is organized as follows: Sect. \ref{MGD} is dedicated  to review the MGD derivation 
as a complete method to deform the Schwarzschild solution and to describe realistic stellar distributions  on finite tension branes. In Sect. \ref{dsfb}, the MGD Yang-Mills-Einstein-Dirac coupled system of field equations, on fluid branes with finite tension, is numerically solved, analyzed and discussed. MGD Yang-Mills-Dirac stellar configurations, whose mass has order of the Chandrasekhar mass, are studied and discussed. For it, the fermionic self-interaction and Yang-Mills coupling constants ranges are physically bounded. Sect. \ref{4} is devoted to conclusions and important perspectives.

\section{The MGD protocol}
\label{MGD}
 The MGD procedure, founded by Ovalle, is constructed to derive high energy scale corrections to GR \cite{ovalle2007,Ovalle:2013xla,darkstars}. 
Fluid branes have a variable tension that emulates cosmological evolution  \cite{Gergely:2008jr,Abdalla:2009pg}. 
The extended MGD derived  the most rigorous brane tension bound, $\sigma \gtrapprox  2.81\times10^{-6} \;{\rm GeV^4}$ \cite{Fernandes-Silva:2019fez}. 

Since the codimension-1 brane is embedded into the bulk, the extrinsic curvature, defined as the Lie derivative of the metric tensor, plays a prominent role in deriving equations of motion. 
As the bulk Riemann tensor can be written in terms of the brane Riemann tensor via the Gauss-Codazzi equation, the effective Einstein's equations on the brane read 
\begin{equation}
\label{5d4d}
{\rm G}_{\mu\nu}
=\Lambda_b g_{\mu\nu}+\mathfrak{T}_{\mu\nu},
\end{equation}  where $8\pi G=1$ is adopted, being $G$ the Newton's constant on the brane, ${\rm G}_{\mu\nu}$ represents the Einstein's tensor and $\Lambda_b$ denotes the brane cosmological running parameter. 
One can cleave the stress-energy tensor in (\ref{5d4d}) into \cite{GCGR} 
\beq
\mathfrak{T}_{\alpha\beta}
=
T_{\alpha\beta}+\mathfrak{E}_{\alpha\beta}+\sigma^{-1} S_{\alpha\beta}
+L_{\alpha\beta}+P_{\alpha\beta}.\label{tmunu}\eeq 
The first term, $T_{\alpha\beta}$, is the stress-energy tensor that represent brane matter and energy, eventually including dark matter and dark energy, whereas the electric part of the bulk Weyl tensor is denoted by $\mathfrak{E}_{\alpha\beta}$. The tensor $\mathfrak{E}_{\alpha\beta}$ is non-local and depends on $\sigma^{-1}$, vanishing in the general-relativistic limit.   Splitting $\mathfrak{E}_{\alpha\beta}$ into traceless transverse ($\mathfrak{E}_{\alpha\beta}^\intercal$) and longitudinal ($\mathfrak{E}_{\alpha\beta}^L$) components, Ref. \cite{GCGR} showed that $\mathfrak{E}_{\alpha\beta}^L$ contains solely brane matter terms. Therefore, the resulting equations of motion are  closed when $\mathfrak{E}_{\alpha\beta}^\intercal=0$, since this term contains Kaluza--Klein bulk gravitons, 
whose backreaction and interaction with brane matter do influence the equations of motion. The $S_{\alpha\beta}$ is a tensor that consists of quadratic terms involving the stress-energy tensor, arising from the extrinsic curvature terms in the projected Einstein tensor. Its intensity is smaller than $\mathfrak{E}_{\alpha\beta}$ \cite{maartens}. In addition, $L_{\alpha\beta}$ encodes the geometry of 
how the brane bents into
the bulk, whereas $P_{\alpha\beta}$ comprises stringy bosonic and fermionic fields in the bulk \cite{Gergely:2008jr,maartens}.

Compact stars are solutions of the system of equations (\ref{5d4d}), with metric 
\begin{equation}\label{abr}
ds^{2} = -A(r) dt^{2} + \frac{1}{B(r)}dr^{2} + r^{2} d\vartheta^2+r^2\sin^2\vartheta d\varphi^2. 
\end{equation} 
One can usually write $A(r)=e^{\upnu(r)}$ and $B(r)=e^{\upchi(r)}$, for a more concise notation in what follows.
The MGD and the MGD-decoupling method, together with their generalizations, can produce analytical solutions of the system (\ref{5d4d}, \ref{tmunu}) \cite{Ovalle:2017fgl,ovalle2007,Casadio:2012rf,darkstars}, including the inner 
solutions concerning  star distributions of radius $R$, given by  \beq
{R} = \frac{\int_0^\infty\,d{\rm r}\, {\rm r}^3\,\varrho({\rm r})}{\int_0^\infty\,d{\rm r}\, {\rm r}^2 \varrho({\rm r})}\eeq  where $\varrho(r)$ represents the density of the star distribution \cite{ovalle2007}. 

Given \be
\label{I}
I
\equiv
\int_0^r\frac{2{\rm r}^2\upnu''({\rm r})+{{\rm r}^2{\upnu'^{2}({\rm r})}}+{4{\rm r}\upnu'({\rm r})}+{4}}
{{{\rm r}^2\upnu'({\rm r})}+{4{\rm r}}}\,d{\rm r}
\,,
\ee
the $g_{rr}$ metric component is then deformed by the embedding of the brane into the bulk \cite{Casadio:2013uma}, 
\begin{eqnarray}
\label{edlrwssg}
e^{-\upchi(r)}
&=&
\mu(r)+\kappa(r)
\ ,
\end{eqnarray}
where \cite{Ovalle:2010zc}
\begin{eqnarray}
\!\!\!\!\!\!\!\!\!\!\kappa(r)\!=\!{e^{-I}\left(\upbeta\!+\!\int_0^r\frac{2{\rm r}e^I}{{{\rm r}\upnu'}+{4}}
\left[L\!+\!\frac{1}{G^2\sigma}\left(\varrho^2\!+\!3p\varrho\right)\right]\right)
d{\rm r}}.
\end{eqnarray}
Besides,  
\begin{eqnarray}
\mu(r) = \begin{cases} 1-\frac{1}{G^2\,r}\int_0^r {\rm r}^2\,\varrho({\rm r}) d{\rm r}
\,,
&
\quad r\,\leq\,R\,,
\\
1-\frac{2\,{GM_0}}{c^2r}
\ ,
&
\quad r>R
\ . 
\end{cases}
\end{eqnarray}
The function $L(r)=L(p(r),\varrho(r),\upnu(r))$ encodes the brane anisotropy induced by gravity in the bulk. 
As the finite brane tension has the strictest bound $\sigma \gtrapprox  2.81\times10^{-6} \;{\rm GeV^4}$ \cite{Fernandes-Silva:2019fez}, the scalar field $\upbeta=\upbeta(\sigma)$ will be  shown to depend on the inverse of $\sigma$, in full compliance with phenomenological data. Hence, the general-relativistic  limit yields  $\lim_{\sigma\to\infty}\upbeta(\sigma)=0$, for outer, $r>R$, solutions. The geometric deformation $\kappa(r)$ in the vacuum ($p=0=\varrho$), denoted by $h^{*}(r)$ in what follows, is minimal, reading \cite{ovalle2007,Casadio:2012rf}
\begin{equation}
\label{def}
h^*(r)
=\upbeta(\sigma)\,e^{-I}
\ .
\end{equation}
Junction conditions match the inner, $r<R$, MGD metric (where $\kappa^*(r)$ is given by Eq. (\ref{edlrwssg}) with $L=0$) 
to the outer solution. The Weyl fluid that circumscribe the brane can be represented by the outer pressure and
the bulk Weyl scalar, respectively given by
\ba
\label{pp2}
\,{\mathfrak{P}^+}{(r)}
&=&
-\frac{\left(1-\frac{4\,{GM}}{3c^2r}\right)\upbeta(\sigma)}{9G^2 \sigma r^3\left(1-\frac{3\,{GM}}{2c^2r}\right)^2},\\
{\mathfrak{U}^+}{(r)}
&=&
\frac{{M}\upbeta(\sigma)}{12G c^2\sigma r^4\left(1-\frac{3\,{GM}}{2c^2r}\right)^2}\ .
\ea
In this case, $p(r)=0=\varrho(r)$ for $r>R$, and the outer solution metric reads \cite{Casadio:2012rf,Ovalle:2016pwp}
\begin{equation}
\label{genericext}ds^2\!=\!-e^{\upnu^+(r)} dt^2\!+\frac{dr^2}{1\!-\!\frac{2GM(\sigma)}{c^2r}\!+\!h^*(r)}+r^{2} (d\vartheta^2+\sin^2\vartheta d\varphi^2).
\end{equation}
Matching conditions at the stellar surface $r=R$, yield \cite{Casadio:2012rf}
\begin{eqnarray}
 \label{ffgeneric1}
{\upnu^-(R)}&=&{\upnu^+(R)}=\log\left(1-\frac{2{GM}}{c^2R}\right),
\\
 \label{ffgeneric2}
\frac{2G}{R} (M-M_0)&=&h^*_R-\kappa^*_R.
\end{eqnarray}
It is worth to emphasize that, in Eq. (\ref{ffgeneric2}), the effective mass is given by $M=M_0+{\cal O}(\sigma^{-1})$, where  terms of order ${\cal O}(\sigma^{-2})$ on are dismissed,  due to the phenomenological lower bound for the brane tension $\sigma \approxeq 2.81\times 10^{-6} \;{\rm GeV^4}$ \cite{Fernandes-Silva:2019fez}. 
The Weyl fluid that bathes the outer stellar configuration implies the matching condition \cite{Ovalle:2013vna}
\be
\label{matchingf}
\!\!\!\!{\sigma}p_R\!+\!\left(\!\frac{\varrho_R^2}{2}\!+\!\varrho_R p_R
\!+\!{2G^4}{}(\mathfrak{U}_R^-\!-\!\mathfrak{U}_R^+)\!\right)
\!+\!{4}{G^4}({\mathfrak{P}_R^-}\!-\!\mathfrak{P}_R^+)
\!=\!
0
\ee
where $\kappa_R^\pm=\lim_{r\to R^\pm} \kappa(r)$.

\bltx{The Schwarzschild-like
solution, $e^{\upnu_{\scalebox{.6}{Sch}}(r)}=e^{-\upchi_{\scalebox{.6}{Sch}}(r)}
=
1-\frac{2\,{GM}}{c^2r}$ can be substituted into Eq.~\eqref{def}, yielding 
\be
\label{defS}
h^*(r)=
-\frac{2\,\upbeta(\sigma)(1-\frac{2GM}{c^2r})}{r\left(r-\frac{3GM}{2c^2}\right)}.
\end{equation}
The function $\upbeta(\sigma)$  can be read off from  Eq.~\eqref{matchingf}, that can be equivalently written as  \cite{Casadio:2012rf}, 
\be
\label{sfgeneric}
R^2p_R+{G^2\kappa^*_R}\left({R\upnu'_R}+1\right)
=
-h^*_R.
\ee
Hence, the outer deformation, 
$h^*(r)$, at $r=R$ has a negative value.
It means that the MGD star event horizon, $r_{\scalebox{.5}{MGD}}=2{GM}/c^2$, is placed nearer the star center than the
standard Schwarzschild event horizon, $r_{\scalebox{.6}{Sch}}=2GM_0/c^2$, since  $M=M_0+{\cal O}(\sigma^{-1})$. One concludes that effects caused by the 
brane embedding into the bulk make gravity to be weaker on the MGD stellar configuration, when one compares it with the  Schwarzschild case \cite{Casadio:2012rf,Ovalle:2007bn}.}

\bltx{Eqs. (\ref{defS}, \ref{sfgeneric}) imply that   \cite{Casadio:2012rf}
\begin{equation}
\label{beta}
\upbeta(\sigma)
=\frac{c^2R-\frac{3\,{GM}}{2}}{c^2R-2GM}
\left[\left({R^2\upnu'_R}{}+R\right)G{\kappa^*_R}+R^3p_R\right].
\end{equation}
One derives $\upbeta(\sigma)$ when the geometrical deformation reads \cite{ovalle2007}
$
\kappa^*(r)
=
\frac{4d\,\iota(r)}{49\,\sigma\,\pi}
y,
$ where $y$ is a numerical value \cite{Casadio:2012rf}, $d=\frac{\sqrt{57}-7}{2R^2}$ and $\iota(r)= (1+dr^2)^{-3}(1+3dr^2)^{-1}$, for $\upnu'(r)=\frac{8dr}{1+dr^2}$. 
The form of $\kappa_R$ yields   \cite{ovalle2007,Casadio:2012rf}  
\begin{equation}
\label{betafinal3}
\upbeta(\sigma)
=\frac{y}{\sigma\,R}\left(\frac{c^2R-\frac{3\,{GM_0}}{2}}{c^2R-{2\,{GM_0}}}\right)\equiv \frac{d_0}{\sigma}.
\end{equation}}

The MGD metric can be then written as  \cite{ovalle2007,Casadio:2012rf}
\begin{subequations}
\ba
\label{nu}
\!\!\!\!\!\!A(r)
&=&e^{\upnu_{\scalebox{.6}{Sch}}(r)}=
1-\frac{2GM}{c^2r}
\ ,
\\
\!\!\!\!\!\!B(r)
&=&
\left[1-\frac{2\,GM}{c^2r}\right]\left(\frac{\mathfrak{l}}{r-\frac{3\,GM}{2c^2}}+1\right)
\ ,
\label{mu}
\ea
\end{subequations} 
where 
\begin{equation}
\label{L}
\bltx{\mathfrak{l}
=
\left(R-\frac{3GM}{2{c^2}}\right)\left(R-\frac{2GM}{c^2}\right)^{\!-1}\!{}\bltx{\frac{d_0}{\sigma}}}.
\end{equation} 
In  the general-relativistic 
limit $\sigma\to\infty$, the Schwarzschild metric is recovered from the MGD metric.

\section{MGD Yang-Mills-Dirac stellar configurations: results, analysis and discussion}
\label{dsfb}
Compact distributions can be described by the MGD. Considering a background fermionic field, $\uppsi$, of mass $m$, MGD stellar configurations can be derived as solutions of Yang-Mills-Einstein-Dirac coupled system field equations. One can take the spin-1/2 fermionic field minimally coupled to gravity (Einstein-Hilbert) and to the SU(2)
 Yang-Mills fields. The action for this system reads \cite{Dzhunushaliev:2019uft}
\begin{equation}
\label{acao3}
	S_{\text{tot}} = S_{\rm EH} +S_\uppsi+S_{\text{YM}},
\end{equation}
where 
 \begin{eqnarray}
S_{\rm EH} &=& - \frac{c^3}{16\pi G}\int d^4 x
		\sqrt{-g} \mathcal{R},\\
S_\uppsi &=& \int d^4 x	\mathcal{L}_\uppsi =	 \int d^4x \left[\frac{i \hbar}{2} \left(
			\bar \uppsi \upgamma^\mu D_\mu\uppsi -
			\bar \uppsi \overleftarrow{D}_{\mu} \upgamma^\mu \uppsi
		\right)\right.\nonumber\\&& \left.\qquad\quad\qquad\quad\qquad\quad- m \bar \uppsi \uppsi + \frac{\uplambda}{2!} \left(\bar\uppsi\uppsi\right)^2\right],
\label{lgian}
\end{eqnarray}
and $
D_\mu \uppsi=  \left(\partial_{\mu} 
+\frac18 \upomega^{a b}_{\;\; \mu}\left[\upgamma^a,  \upgamma^b\right]-  \frac{i{\tt{g}}}{2}
		A^a_\mu \sigma^a\right)\uppsi$, 
where $\upomega_{a b \mu}$ denotes  the spin-connection.
The set $\{\upgamma^\mu\}$ is constituted by gamma matrices, satisfying  $\{\upgamma^\mu, \upgamma^\nu\}=2g^{\mu\nu}\mathbb{I}$, and $\bar\uppsi=\uppsi^\dagger\upgamma_0$ is the spinor conjugate. Dirac matrices in curved spaces,
$\upgamma^a = e^{a}_{\;\mu}\upgamma^\mu$, are computed when one employs tetrads
$e^{a}_{\;\mu}$.  The term $i \frac{{\tt{g}}}{2}  A^a_\mu \sigma^a\uppsi$ describes the coupling between the spin-1/2 fermionic field and the Yang-Mills field,
where ${\tt{g}}$ is the SU(2) coupling constant and $\sigma^a$ denote the set of Pauli matrices.
Besides, ($a,b,c=1,2,3$)
\begin{equation}
\label{lym}
S_{\text{YM}} = \int d^4x\mathcal{L}_{\text{YM}}= -\int d^4x
		\frac{1}{4} F^c_{\mu\nu}F^{c \mu\nu},
\end{equation}
where $
	F^c_{\mu \nu} = \partial_{[\mu} A^c_{\nu]} +
	{\tt{g}} \epsilon^c_{\;a b } A^a_\mu A^b_\nu
$ is the Yang-Mills tensor field strength and $A^a_{\mu}$ represent the Yang-Mills gauge potential.

With the action (\ref{acao3}) and the MGD, one derives the MGD Yang-Mills-Einstein-Dirac system of equations of motion, with MGD-decoupling
 \begin{subequations}
\begin{eqnarray}
	\!\!\!\!\!\!\!\!\!\!\!\!{\rm G}_{\mu\nu} -
	\Lambda_b g_{\mu\nu}- (1+\upzeta)\mathcal{T}_{\mu\nu} &=&\mathfrak{T}_{\mu\nu} ,
\label{eins} \\
\!\!\!\!\!\!\!\!\!\!\!\!	\left[i \hbar \upgamma^\mu D_\mu - mc\mathbb{I} + \uplambda(\mathbb{I}  -\bar\uppsi\uppsi)\right]\uppsi&=& 0,
\label{einssp}\\
\!\!\!\!\!\!\!\!\!\!\!\!	\bar\uppsi \left[i \hbar \overleftarrow{D}_{\mu}  \upgamma^\mu + mc\mathbb{I} + \uplambda(\mathbb{I}  -\bar\uppsi\uppsi)\right]&=& 0,
\label{einssp1}\\
\!\!\!\!\!\!\!\!\!\!\!\!\!\!\!\!\!\!\!\!\frac{1}{\sqrt{-g}}\partial_\nu\left(\sqrt{-g}F^{a\mu\nu}\right)\!+\!{\tt{g}} \epsilon^{a}_{\;bc} A^b_\nu F^{c\mu\nu}
\!&\!=\!&\!\frac{{\tt{g}}\hbar c}{2}\bar\uppsi\upgamma^\mu\sigma^a\uppsi,
\label{YM}
\end{eqnarray}
\end{subequations} where  \bltx{$\upzeta(\sigma)\simeq\sigma^{-1}$} governs the 
 MGD decoupling. The spin-1/2 fermionic stress-energy tensor is given by  
\begin{eqnarray}
	\mathcal{T}_{\mu\nu} &=&{i\hbar}g_{\nu}^{\;\,\tau}\bar\uppsi \left(\upgamma_{(\mu} D_{\tau)}+\overleftarrow{D}_{(\mu}\upgamma_{\tau)}\right)\uppsi
-2\delta_{\mu\nu}{\uplambda} \left(\bar\uppsi\uppsi\right)^2\nonumber\\&&-4F^a_{\mu\tau}F^{a\tau}_{\nu}+\delta_{\mu\nu} F^a_{\alpha\beta}F^{a\alpha\beta}. 
\label{stensor}
\end{eqnarray}   Besides, due to the tiny phenomenological bound for the term $\mathfrak{l}$ in Eq. (\ref{L}) \cite{Casadio:2015jva}, that compounds the MGD metric component (\ref{mu}), the off-diagonal components of \eqref{stensor} are equal to zero.

The Yang-Mills field has the usual form, emulating a non-Abelian monopole,
\begin{eqnarray}
	A^c_i &=&  \frac{1 - f(r)}{\tt g} 
	\begin{pmatrix}
		 0 & \sin \varphi &  \frac{\sin(2\vartheta)}{2} \cos \varphi \\
		 0 & -\cos \varphi &  \frac{\sin(2\vartheta)}{2} \sin \varphi \\
		 0 & 0 & - \sin^2 \vartheta
	\end{pmatrix},
\label{ymi}
\end{eqnarray}
being its temporal component equal to zero. 

The following two spinor ans\"atze\begin{subequations} \beq\label{spn1}
\!\!\!\!\!\!\!\!\!\!\!\!\!\!\!\!\!\!\!\!\uppsi_1(r,t)&=&{2} \exp\left({-i \frac{E t}{\hbar}}\right) \left(\begin{matrix}
			0 \\ \alpha_1(r)\\i \alpha_2(r) \sin \vartheta e^{- i \varphi}\\- i\alpha_2(r) \cos \vartheta\end{matrix}\right),\\
\!\!\!\!\!\!\!\!\!\!\!\!\!\!\!\!\!\!\!\!	\uppsi_2(r,t)&=&-{2} \exp\left({-i \frac{E t}{\hbar}}\right) \left(\begin{matrix}			\alpha_1(r) \\0\\
		 i\alpha_2(r) \cos \vartheta \\  i\alpha_2(r) \sin \vartheta e^{i\varphi} \\
		\end{matrix}\right),
	\label{spinor}
\eeq
\end{subequations}
where   $\alpha_1,\alpha_2:[0,\infty)\to\mathbb{R}$, were proposed  to study coupled systems \cite{Herdeiro:2017fhv,Dzhunushaliev:2018jhj}.

Substituting \eqref{spn1} and  \eqref{spinor}  in the MGD Yang-Mills-Einstein-Dirac system of equations of motion (\ref{eins}) -- (\ref{YM}), the coupled system of ODEs is obtained, in terms of the MGD metric (\ref{abr}). For the analysis that follows, one denotes the dimensionless radius, $x =  \frac{m r}{\hbar}$ and  $\xi=\frac{\pi m}{\hbar c{\tt g}^2 M_p^2}$ (and by $(\;\;)^\prime$ the derivative with respect to $x$), being $M_p =\sqrt{\hbar c/G}\approxeq 1.2209 \times 10^{22} {\rm MeV}/c^2=2.176 \times 10^{-8}$ Kg  
 the Planck mass, and the following rescaled quantities   \cite{Dzhunushaliev:2019uft,Dzhunushaliev:2018jhj}:
\begin{subequations}
\beq
\label{mud}
		&&
	E\mapsto\tilde E = \frac{E}{m c^2}, \qquad\quad
	{\tt g}\mapsto   \tilde {\tt g}={\tt g}\sqrt{\hbar c},\\
	&&M \mapsto \tilde{M}=\frac{mM}{M_{p}^2},\quad \qquad
	\lambda\mapsto\tilde \lambda =\frac{\lambda m^2c}{4\tilde{\tt g}^2 \hbar^3}, \\
&& \alpha_{1,2} \mapsto \tilde{\alpha}_{1,2}=2\left(\frac{mc}{\hbar}\right)^{\frac32} \tilde{\tt g} \alpha_{1,2}.
\eeq
Hence, the  MGD Yang-Mills-Einstein-Dirac coupled system of ODEs reads 
\end{subequations}
\begin{widetext}
\begin{subequations}
\begin{eqnarray}
	\!\!\!\!\!\!\!\!\!\!\!\!\!\!\!\!\!\!\!\!\tilde \alpha_2^\prime(x) \!+\! \left[
		\frac{A^\prime(x)}{4\sqrt{A(x)B(x)}} \!+\!\frac{1}{x}\right] \tilde \alpha_2(x) \!+\!\frac{f(x) \tilde \alpha_2(x)}{x\sqrt{B(x)}}
\!+\! \left[
		\frac{1}{\sqrt{A(x)}} \!+\!8 \lambda\,\frac{\tilde \alpha_2^2(x) \!-\! \tilde \alpha_1^2(x)}{\sqrt{B(x)}}\!-\! \frac{\tilde E}{\sqrt{A(x)B(x)}}
	\right]\tilde \alpha_1(x)&=& 0,
\label{fe1}\\
\!\!\!\!\!\!\!\!\!\!\!\!\!\!\!\!\!\!\!\!\tilde \alpha_1^\prime(x) \!+\! \left[
		\frac{A^\prime(x)}{4\sqrt{A(x)B(x)}} \!+\! \frac{1}{x}\right] \tilde \alpha_1(x) \!-\!\frac{f(x) \tilde \alpha_1(x)}{x\sqrt{B(x)}}
\!+\! \left[
		\frac{1}{\sqrt{A(x)}} \!+\!8\tilde \lambda\,\frac{\tilde \alpha_2^2(x) \!-\! \tilde \alpha_1^2(x)}{\sqrt{B(x)}}\!+\! \frac{\tilde E}{\sqrt{A(x)B(x)}}
	\right]\tilde \alpha_2(x)&=& 0,
\label{fe2}\\
\!\!\!\!\!\!\!\!\!\!\!\!\!\!\!	\!\!\!\!\!\!\!\!\!\!\!\!\!\!\! \tilde  2\xi x^2\left[\frac{\left(f^2(x)\!-\!1\right)^2}{x^4}\!+\!\frac{2  \sqrt{A(x)B(x)} f^{\prime 2}(x)}{x^2}\!+\!
\frac{4\tilde E (1\!+\!\upzeta)}{\sqrt{B(x)}}\left(\tilde \alpha_1^2(x)\!+\!\tilde \alpha_2^2(x)\right)\!+\!16 \lambda\left(\tilde \alpha_2^2(x)\!-\!\tilde \alpha_1^2(x)\right)^2
\right]&=&M^\prime(x),
\label{fe3}\\
\!\!\!\!\!\!\!\!\!\!\!\!\!\!\!\!\!\!\!\!\!\!\!\!\!\!\!\!\!\!\frac{8\xi x}{\sqrt{B(x)}}\left[
\frac{\tilde E (1\!+\!\upzeta)}{\sqrt{A(x)B(x)}}\left(\tilde \alpha_1^2(x)\!+\!\tilde \alpha_2^2(x)\right)\!+\! \tilde \alpha_1(x) \tilde \alpha_2^\prime(x)\!-\!\tilde \alpha_2(x) \tilde \alpha_1^\prime(x)\!+\!\frac{\sqrt{B(x)}f^{\prime 2}(x)}{x^2}
\right]&=&0,
\label{fe4}\\
\!\!\!\!\!\!\!\!\!\!\!\!\!\!\!\!\!\!\!\!\!\!\!\!\!\!\!\!\!\! f^{\prime\prime}(x)\sqrt{A(x)B(x)}\!+\!{A^\prime(x)} \!+\!\frac{f(x)}{x^2}\left(1\!-\!f^2(x)\right)-2x{\tilde \alpha_1(x)\tilde \alpha_2(x)}&=&0.
\label{fe5}
\end{eqnarray}
\end{subequations}
\end{widetext}
As Planckian compact stellar distributions are not aimed to be studied in this work, 
the fermion mass is considered to be negligible when compared to the Chandrasekhar mass ${M_{p}^3}{m^{-2}}$.
The SU(2) Yang-Mills running constant  ${\tt g}$ is usually inversely proportional to the energy scale of the system. Since for the energy scale of compact stellar distributions that interest us has the coupling constant {\tt g} of the order of unity, then $\tilde{\lambda}$ must be also of this order  \cite{Dzhunushaliev:2019uft}. 

The MGD of the Schwarzschild metric, (\ref{abr}), having temporal and radial  coefficients (\ref{nu}, \ref{mu}), satisfies separately the effective Einstein field equations on the brane \eqref{eins}, with MGD-decoupling. Now we must verify whether the MGD metric and the ans\"atze (\ref{spn1}, \ref{spinor}), together with the Yang-Mills field (\ref{ymi}), do satisfy the whole Yang-Mills-Einstein-Dirac coupled system (\ref{fe1}) -- (\ref{fe5}). For it, there will be constraints in the form of the 
spinor fields coefficients, $\alpha_1(r)$ and $\alpha_2(r)$, in (\ref{spn1}, \ref{spinor}). 
Integrating the system (\ref{fe1}) -- (\ref{fe5})  numerically, one can use boundary conditions with respect to the center of compact distribution \cite{Dzhunushaliev:2019uft}:
\beq
	&&\tilde \alpha_1(x)\approxeq \alpha_{1c} + \frac{1}{2!}\upalpha_1 x^2, \quad\qquad
	\tilde \alpha_2(x)\approxeq \tilde \upalpha_2 x,\label{exps0}\\
	&&\tilde M(x)\approx \frac{1}{3!}\tilde M_3 x^3, \quad\qquad f(x) \approxeq 1 + \frac{1}{2!} f_2 x^2.
\label{exps}
\eeq
The parameter $\alpha_{1c}$ indicates $\lim_{x\to 0}\alpha_1(x)$. The coefficients $\upalpha_1$, $\upalpha_2$ and $M_3$  in Eqs. (\ref{exps0}, \ref{exps}) are obtained by the solution and integration of the system \eqref{fe1} - \eqref{fe5}, whereas $f_2$,  $\tilde E$  and $\alpha_{1c}$ are completely arbitrary parameters, but, of course, the values of physical interest must generate regular compact solutions in the asymptotically flat limits $\lim_{x\to \infty}A(x)= 1= \lim_{x\to \infty}B(x)$ and $\lim_{x\to \infty} f(x) =\pm 1$ \cite{Dzhunushaliev:2019uft}. 
The ADM mass of the stellar configuration corresponds to $M_\infty=\lim_{x\to \infty}\tilde{M} = \frac{mM}{M_p^2}$.

Refs. \cite{Dzhunushaliev:2018jhj,Dzhunushaliev:2019kiy} showed that 
one can derive stellar configurations with Chandrasekhar mass order for $\tilde \lambda<0$, such that  $|\tilde \lambda|\gg \xi $. It also happens for the MGD Yang-Mills-Dirac stars, where one can take $\tilde \lambda\approx 1$ and small values of $\xi$. Otherwise, choosing $\xi \approx 1$ implies Planckian stars.
Typical masses, $\sim M^2_p/m$ and radii, $\sim m^{-1}$, are regarded for the  stellar configurations here studied. For fermion mass $m\approx 1$ GeV, corresponding to nucleons, the MGD Yang-Mills-Dirac star mass equals $\sim10^{10}$ Kg, being much smaller than the Chandrasekhar mass $\sim M^3_p/m^2\gg M_\odot$, where $M_\odot = 1.989 \times 10^{30}$ Kg denotes the Solar mass. On the other hand, for $m \approx 10^{-10}$ eV, the MGD Yang-Mills-Dirac star mass equals $\approx M_\odot$ and its radius $\approx 10$ Km, being 
feasible for eventual gravitational waves observations.
\begin{figure}[H]
\centering\includegraphics[width=7.cm]{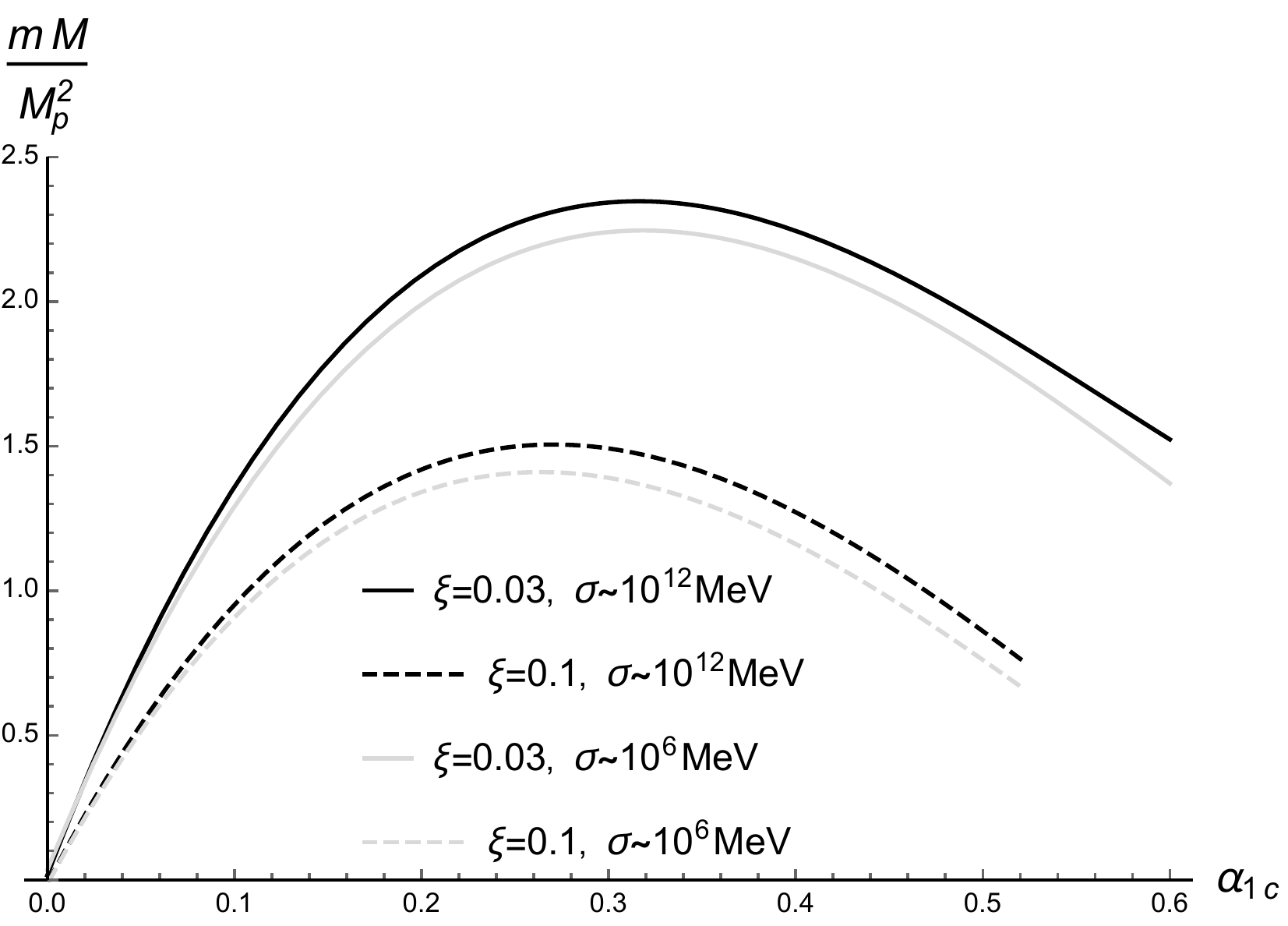}
\caption{ADM mass of MGD Yang-Mills-Dirac stars, as a function of the central value of the spinor field $\alpha_1$ component, $\alpha_{1c}$, for two values $\xi=0.03$ and $\xi= 0.1$, and $\tilde\lambda=-1$.
 Two values of the brane tension, $\sigma \approx 2.8\times 10^{6} \;{\rm MeV^4}$ and $\sigma \sim 10^{12} \;{\rm MeV^4}$ are analyzed, for the MGD-decoupling parameter value $\upzeta=0.1$.}
\label{fi1}
\end{figure} 
\begin{figure}[H]
\centering\includegraphics[width=8.5cm]{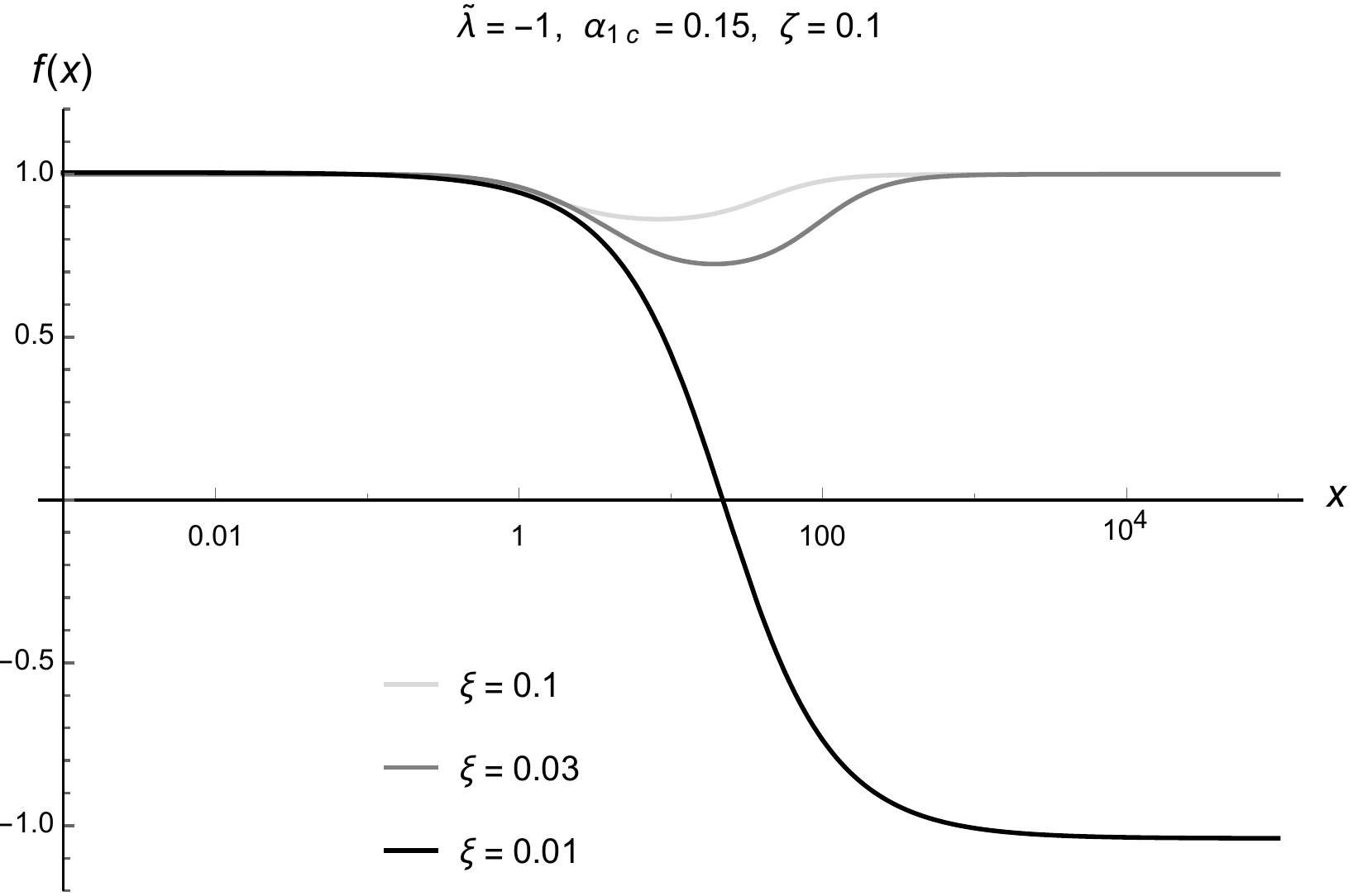}
\caption{Linear-log plot of the Yang-Mills field, $f(x)$, as a function of dimensionless stellar radius, $x$, for $\xi=0.01$, $\xi=0.03$ and $\xi=0.1$, being the central value of the spinor field $\alpha_1$ component, $\alpha_{1c}=0.15$, adopted, for $\tilde\lambda=-1$.
The brane tension $\sigma \sim 10^{12} \;{\rm MeV^4}$ is taken and $\upzeta=0.1$ rules the MGD-decoupling.}
\label{fi2}
\end{figure}

After numerical integration, the MGD Yang-Mills-Dirac compact stellar configuration ADM mass parameter, $\tilde{M}_\infty = \frac{m M}{M_{\rm p}^2}$, as a function of the central value $\alpha_{1c}$,  has the profile illustrated in Fig.~\ref{fi1}, for the MGD-decoupling parameter $\upzeta=0.1$. Two different values of the Yang-Mills coupling constant are considered, as well as two different values of the finite brane tension. In fact, as the most precise current bound on the variable brane tension is $\sigma \gtrapprox  2.81\times10^6 \;{\rm MeV^4}$ \cite{Fernandes-Silva:2019fez}, then Fig. \ref{fi1} 
 takes this lower brane tension limit, and the distinct case  $\sigma \sim 10^{12} \;{\rm MeV^4}$, to 
study the physical differences among these cases. It is worth to emphasize that the general-relativistic limit corresponds to a rigid brane, making $\sigma\to\infty$ and \bltx{$\upzeta\to0$}. 

 In Fig.~\ref{fi1}, each plot has a peak in the MGD Yang-Mills-Dirac stellar mass, at some value of
central value $\alpha_{1c}$ of the spinor field component. For the brane tension $\sigma \sim 10^{12} \;{\rm MeV^4}$ and $\xi=0.03$  the maximal mass equals  $2.32 \tilde{M}_\infty$ at $\tilde{\alpha}_{1c}=0.318$,  whereas for $\xi=0.1$ the maximal mass is equal to $1.50 \tilde{M}_\infty$, for $\tilde{\alpha}_{1c}=0.279$. On the other hand, for  the most strict bound $\sigma \approx  2.81\times10^6 \;{\rm MeV^4}$,
when $\xi=0.03$  the maximal mass equals $2.22 \tilde{M}_\infty$ at $\tilde{\alpha}_{1c}=0.332$,  whereas for $\xi=0.1$ the maximal mass is $1.40 \tilde{M}_\infty$, for $\tilde{\alpha}_{1c}=0.270$. 
One realizes that the bigger the brane tension, the bigger the maximal mass is, at smaller values of $\tilde{\alpha}_{1c}$. MGD Dirac stars were studied in a similar context, where a maximal mass was identified to a transition point, splitting stable and unstable MGD Dirac compact stellar configurations \cite{daRocha:2020rda}. 

Fig.~\ref{fi1} is motivated by the sign of the binding energy,
which is defined as the difference between the energy of ${\rm n}_f$ free particles and the total energy of the system. The number
 ${\rm n}_f$ corresponds to a Noether charge, being computed when the 4-current density $j^\rho=\sqrt{-g}\bar \uppsi \upgamma^\rho \uppsi$ is taken into account 
as 
$
{\rm n}_f=\int_{\mathbb{R}^3} j^0 d^3 x
$ \cite{Dzhunushaliev:2019uft}, 
where $j^0 = r^2 \sin{\vartheta} \frac{\left(\uppsi^\dag \uppsi\right)}{\sqrt{A}}$.
In dimensionless variables, 
\begin{equation}
{\rm n}_f=\frac{8\xi M_p^2}{m^2}\int_0^\infty \frac{ \tilde\alpha_1^2(x)+\tilde\alpha_2^2(x)}{\sqrt{A(x)}}x^2 dx.
 \label{p0}
\end{equation}
As in the temporal component (\ref{nu}) of the MGD metric the denominator equals the Schwarzschild one, the particle number in Eq. (\ref{p0}) is the same as in the Schwarzschild metric.
As accomplished in Ref. \cite{Dzhunushaliev:2019uft}, stellar distributions having negative binding energy are unstable. Hence, the plots in Fig.~\ref{fi1}
 have the range of the variable $\alpha_{1c}$ compatible to positive values of the binding energy.
 The higher the value of $\xi$, the narrower the range of the variable $\alpha_{1c}$ is. 

Fig.~\ref{fi1} also illustrates how the maximal MGD Yang-Mills-Dirac stellar mass
 increases as a function of the central value, $\alpha_{1c}$, of the spinor component $\alpha_1$. For decrements of $\xi$, the maxima of the total ADM mass  $M^{max}$ increase.
It is worth to emphasize that  Fig.~\ref{fi1} shows the dependence of
 $M^{max}$ on $\xi$, for small values of $\xi$: 
  \begin{equation}
\label{maxx}
	M^{max}\approx \frac{0.403(1+\upzeta)}{\sqrt{\xi}}\frac{M_p^2}{m}.
\end{equation}

Outer solutions for the Yang-Mills field $f$, with respect to the dimensionless radius, $x$, of the stellar distributions are shown in Figs.~\ref{fi2}. Three values $\xi=0.01$, $\xi=0.03$ and $\xi=0.1$ are employed in the analysis, for  $\alpha_{1c}=0.15$ and $\tilde\lambda=-1$. The brane tension value $\sigma \sim 10^{12} \;{\rm MeV^4}$ is adopted and $\upzeta=0.1$ governs the MGD-decoupling.
 Fig. \ref{fi2} illustrates the fine dependence of the Yang-Mills field $f$ on the $\xi$ parameter.

\begin{figure}[H]
\centering\includegraphics[width=8.6cm]{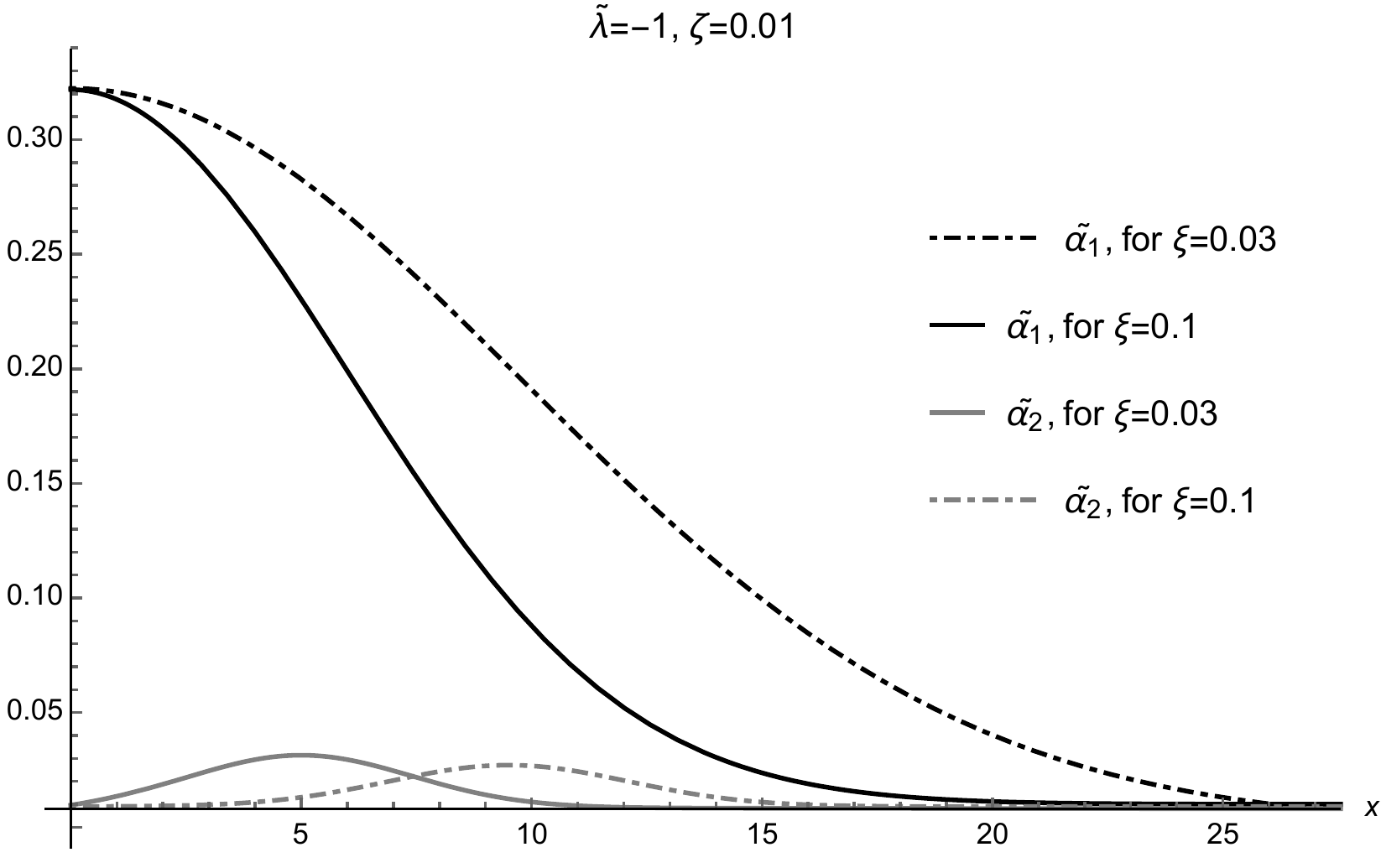}
\caption{Spinor field components, $\alpha_1(x)$ and $\alpha_2(x)$ as a function of dimensionless radius~$x$ for $\xi=0.01$, $\xi=0.03$ and $\xi=0.1$, being the central value  $\alpha_{1c}=0.15$ adopted, for $\tilde\lambda=-1$.
The brane tension $\sigma \sim 10^{12} \;{\rm MeV^4}$ is taken and $\upzeta=0.01$ rules the MGD-decoupling.}
\label{fi3}
\end{figure}

\begin{figure}[H]
\centering\includegraphics[width=8.6cm]{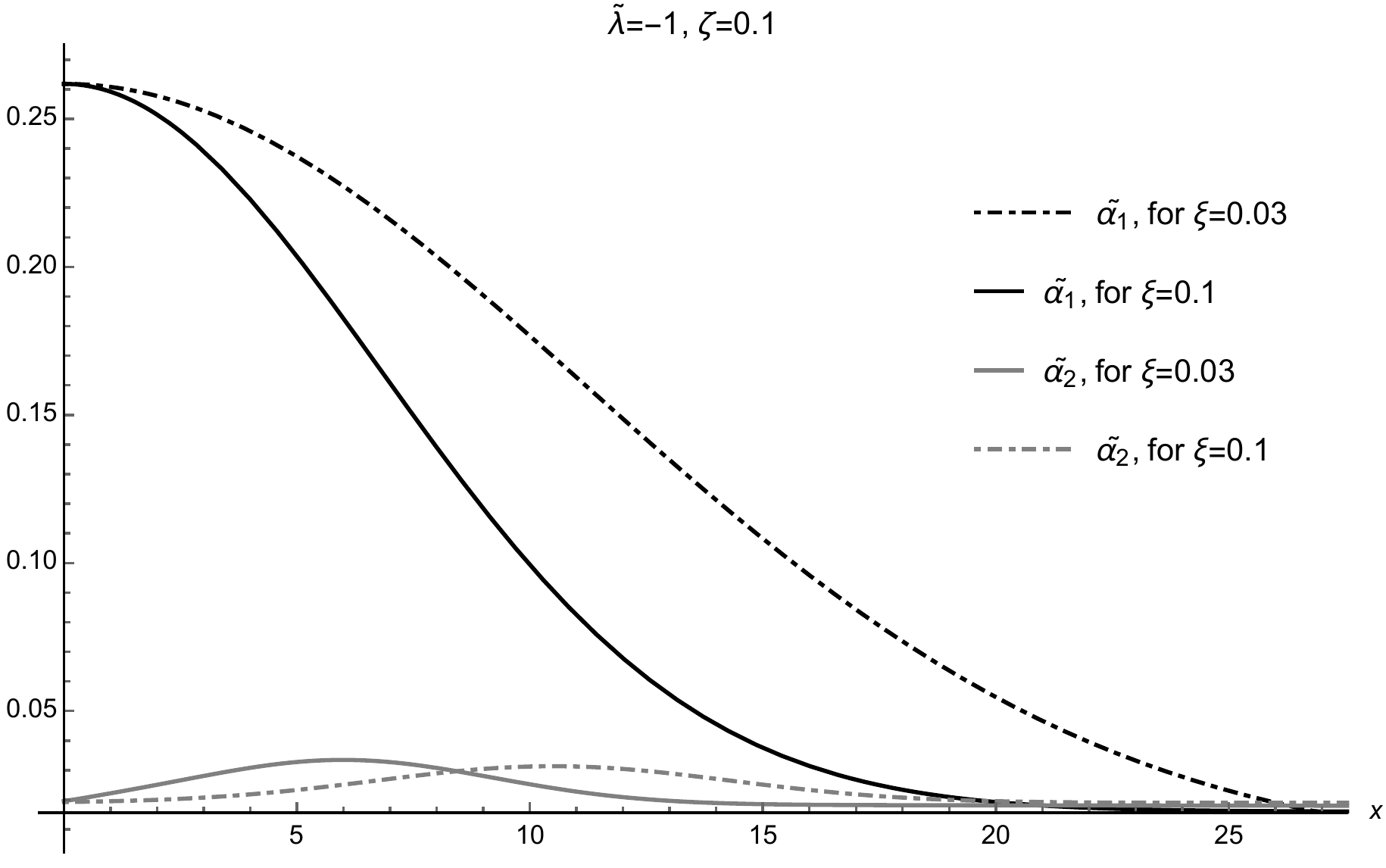}
\caption{Spinor field components, $\alpha_1(x)$ and $\alpha_2(x)$ as a function of dimensionless radius~$x$ for $\xi=0.01$, $\xi=0.03$ and $\xi=0.1$, where $\alpha_{1c}=0.15$ adopted, for $\tilde\lambda=-1$.
The brane tension $\sigma \sim 10^{12} \;{\rm MeV^4}$ is taken and $\upzeta=0.1$ rules the MGD-decoupling.}
\label{fi4}
\end{figure}

The spinor field components $\alpha_1(x)$ and $\alpha_2(x)$, as a function of dimensionless radius~$x$, are illustrated in Figs. \ref{fi3} and \ref{fi4}, respectively for the MGD-decoupling parameter $\upzeta = 0.01$ and $\upzeta = 0.1$ and two different values  $\xi=0.03$ and $\xi=0.1$, being the central value of the spinor field $\alpha_1$ component, $\alpha_{1c}=0.15$ adopted, for $\tilde\lambda=-1$. Again, the brane tension value $\sigma \sim 10^{12} \;{\rm MeV^4}$ is chosen. 
The SU(2) Yang-Mills field does not induce significant modifications on the stellar configurations, in what concerns its coupling to the fermionic fields. Therefore, qualitatively the spinor fields
profiles in Figs. \ref{fi3} and \ref{fi4} do not significantly change, when compared to the MGD Dirac stellar distributions in Ref.~\cite{daRocha:2020rda} and their GR limit \cite{Dzhunushaliev:2018jhj}.


The $\xi\ll 1$ regime can be further studied. For these values, there is a relation between the  spinor field components, asserting that  $\tilde \alpha_2\ll\tilde \alpha_1$, as seen in Figs. \ref{fi3} and \ref{fi4}. Hence, Eq.~\eqref{fe1} yields, in this regime, 
\begin{equation}
 \label{u_approx}
 \tilde \alpha_{1\star} = \frac{1}{2\sqrt{2}}\sqrt{
 -\left(1-\frac{\tilde E}{\sqrt{B}}\right)},
\end{equation}
where  $\tilde \alpha_{1\star}=\sqrt{|\tilde \lambda|}\tilde \alpha_{1}$ \cite{Dzhunushaliev:2019kiy}.
Substituting Eq. (\ref{u_approx}) into Eqs.~(\ref{fe3}, \ref{fe4}), the notation 
$x_\star=\sqrt{\frac{\xi}{|\tilde \lambda|}} x$ and $\tilde M_\star=\sqrt{\frac{\xi}{|\tilde \lambda|}} \tilde M$
yields 
\begin{eqnarray}
	\frac{d \tilde M_\star}{d x_\star} &=& 8 x_\star^2 \left(
\frac{\tilde{E}(1+\upzeta)}{\sqrt{A}}-4 \tilde \alpha_{1\star}^2
	\right)\tilde  \alpha_{1\star}^2,
\label{fe3a}
\end{eqnarray}

\begin{figure}[H]
\centering\includegraphics[width=7.6cm]{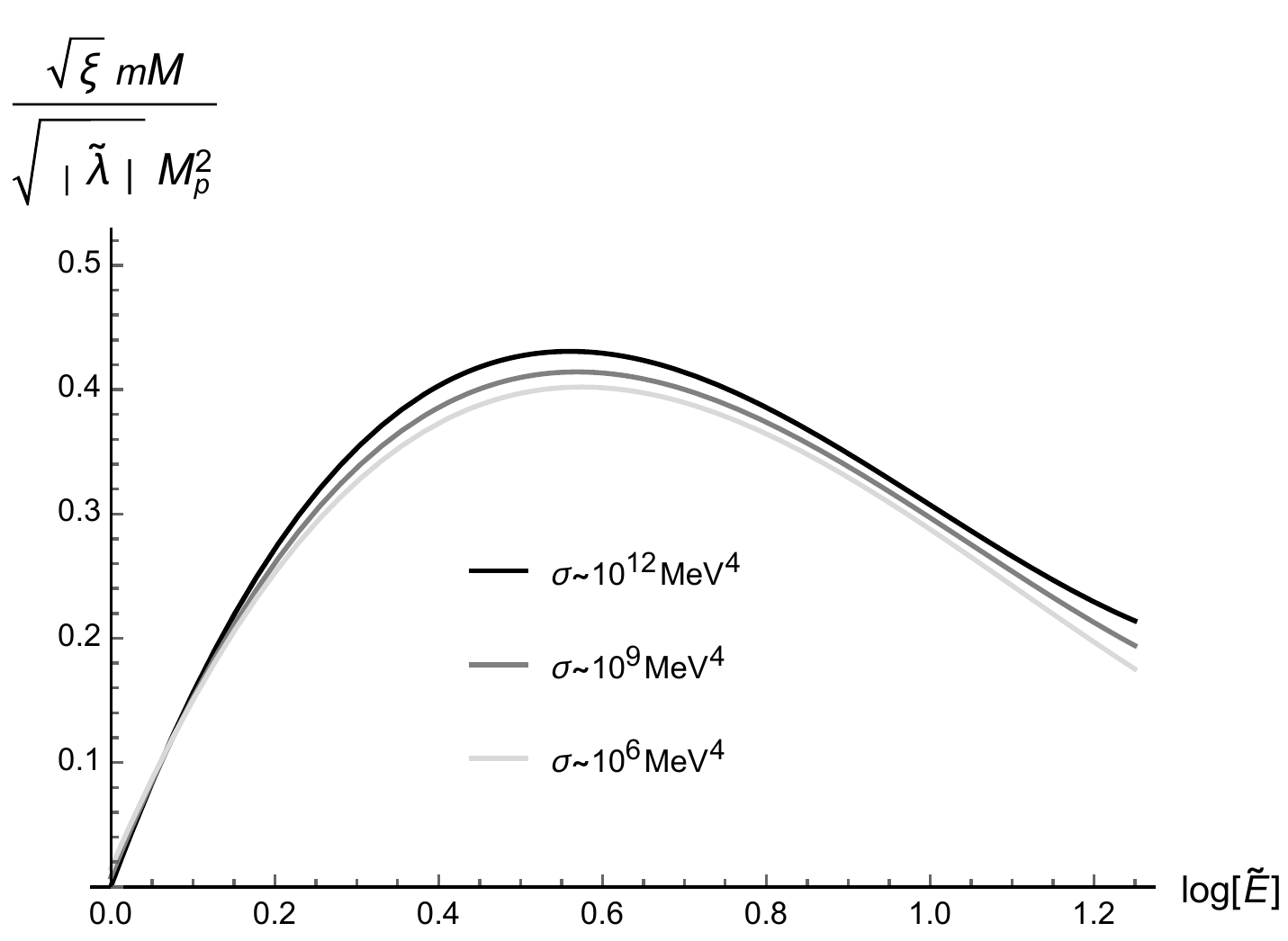}
\caption{Dimensionless total ADM mass $M_{\star\infty}$ of MGD Yang-Mills-Dirac stellar configurations as a function of the energy of the stationary part of the spinor fields, $\tilde{E}$.  Brane tension values $\sigma \sim 10^{12} \;{\rm MeV^4}$ (black line), $\sigma \sim 10^{9} \;{\rm MeV^4}$ (gray line) and $\sigma \sim 10^{6} \;{\rm MeV^4}$ (light gray line) are taken and $\upzeta=0.1$ rules the MGD-decoupling.}
\label{fi5}
\end{figure}

Since Eq. (\ref{fe3a}) lacks the explicit appearance of the parameter $\xi$, one can employ Eq. (\ref{fe3a})  to determine the ADM (rescaled) mass
$\tilde M_{\star\infty} = \lim_{x\to\infty}M_\star = \sqrt{\frac{\xi}{|\tilde \lambda|}}\frac{M_{\star}m}{M_p^2}$, with respect to $\tilde E$.
 Fig.~\ref{fi5} represents the numerical solutions, for three different values of the brane tension.
 For each value of the brane tension, the maximal mass of the MGD Yang-Mills-Dirac stellar configuration is given by
\begin{equation}
\label{mstar}
	M_\star^{max} \approx 0.403(1+\upzeta)\sqrt{\frac{|\tilde \lambda|}{\xi}}  \frac{M_p^2}{m}.
\end{equation}
It is worth to emphasize that when $\tilde\lambda=-1$, Eq. (\ref{maxx}) is immediately recovered.
Besides, our results lead to the ones in the general-relativistic limit in Ref. \cite{Dzhunushaliev:2019uft}.
The brane tension values $\sigma \sim 10^{12} \;{\rm MeV^4}$ (black line) is nearer the general-relativistic limit, whereas $\sigma \approxeq 2.81\times 10^{6} \;{\rm MeV^4}$ represents the  phenomenological current bound \cite{Fernandes-Silva:2019fez}, being more realistic for astrophysical applications. When $\sigma \sim 10^{12} \;{\rm MeV^4}$, the maximal mass reads $M^{max}_{\star\infty}=0.436$, at $\log(\tilde E)=0.56$; for $\sigma \sim 10^{9} \;{\rm MeV^4}$, one has $M^{max}_{\star\infty}=0.419$, at $\log(\tilde E)=0.54$; and when $\sigma \approxeq 2.81\times 10^{6} \;{\rm MeV^4}$, the maximal mass is given by $M^{max}_{\star\infty}=0.402$, at $\log(\tilde E)=0.59$. The bigger the brane tension, the bigger the peak corresponding to the stellar maximal mass is. 

As accomplished to the ADM mass of the MGD Yang-Mills-Dirac stellar configurations,  their effective radius can be also studied. For it, an analog approach will be here employed as the one in the GR limit in Refs. \cite{Jetzer:1989av,Dzhunushaliev:2019uft}, defining the effective radius of MGD Yang-Mills-Dirac stellar configurations as 
\begin{equation}
\!\!\!\!\!{\rm R}\!=\!\frac{1}{{\rm n}_f}\!\!\int_{\mathbb{R}^3} \!\!\!r\, j^0 r^2\sin\vartheta dr d\Omega \!=\! \frac{\hbar}{{\rm n}_f mc}\int_0^\infty \frac{\tilde \alpha_1^2\!+\!\tilde \alpha_2^2}{\sqrt{A}}x^3 dx.
\label{raio}
\end{equation}
It  depends on the parameter $\xi$, in the regime $\xi\ll 1$, as
\begin{equation}\label{rmax}
 {\rm R}_\star^{max}\approx 1.078(1+\upzeta) \sqrt{\frac{|\tilde \lambda|}{\xi}}\frac{\hbar}{mc}.
\end{equation}
This result leads to the ones in the general-relativistic limit in Ref. \cite{Dzhunushaliev:2019uft}.

\section{Conclusions and prospects}
\label{4}
MGD Yang-Mills-Dirac stellar configurations, consisting of non-linear spinor fields minimally coupled to gravity in a SU(2) Yang-Mills background, on a fluid brane with finite tension, were studied and discussed. Two feasible fermionic fields ans\"atze were employed, as compatible spinor solutions to the MGD metric. 
These spinor fields that compose the MGD Yang-Mills-Dirac stars, have components plot in Figs. \ref{fi3} and \ref{fi4}, respectively for the MGD-decoupling parameter $\upzeta = 0.01$ and $\upzeta = 0.1$ and two different values  $\xi=0.03$ and $\xi=0.1$, for a specific value of the central value, $\alpha_{1c}=0.15$ chosen. The spinor self-interaction parameter $\tilde\lambda=-1$ was adopted throughout 
the analysis. In fact, for these values and small values of $\xi$, stellar configurations with mass of the order of the Chandrasekhar mass can be derived, also avoiding Planckian  compact MGD Yang-Mills-Dirac stellar configurations. 
The SU(2) Yang-Mills field was shown not to have a significant influence on the stellar configurations, mainly in what regards the coupling to the fermionic fields. However, Fig. \ref{fi2} reveals a fine dependence of the Yang-Mills field $f$ on the $\xi$ parameter. 
Hence, the spinor fields
profiles in Figs. \ref{fi3} and \ref{fi4} do not significantly change, when compared to the MGD Dirac stellar distributions in Ref.~\cite{daRocha:2020rda} and to GR limit in Ref. \cite{Dzhunushaliev:2019uft} as well. 
Besides, Fig. \ref{fi5} showed the total ADM mass of MGD Yang-Mills-Dirac stellar configurations as a function of the energy of the stationary part of the spinor fields.   The lower the brane tension, the lower the peak corresponding to the MGD Yang-Mills-Dirac stellar maximal mass is.

The MGD Yang-Mills-Dirac stars are qualitatively similar to their general-relativistic 
limit counterparts. In fact, the MGD parameter, $\mathfrak{l}$, in Eq. (\ref{L}) attains small values  \cite{Casadio:2015jva}. However, even small, it does affect the solutions found by solving the system (\ref{fe1} -- \ref{fe5}). Besides, the MGD-decoupling parameter $\upzeta\sim\sigma^{-1}$ also affects, for instance,  
the results in (\ref{eins}, \ref{fe3}, \ref{fe4}, \ref{maxx}, \ref{fe3a}), making the maximal mass and the maximal radius, respectively in Eqs. (\ref{mstar}, \ref{rmax}), to vary, according to the  most strict bound for the brane tension $\sigma \gtrapprox  2.81\times10^6 \;{\rm MeV^4}$ \cite{Fernandes-Silva:2019fez}. 
 In the general-relativistic limit, when  \bltx{$\sigma\to\infty$}, and consequently $\upzeta\to0$, the GR results in Ref. \cite{Dzhunushaliev:2019uft} are recovered. 
 
Although relatively similar, from the qualitative point of view, to some results in the GR limit of Ref. \cite{Dzhunushaliev:2019uft}, the MGD Yang-Mills-Dirac stars stellar configurations here studied bring a more realistic model, being more feasible, whose eventual gravitational wave observations in LIGO and eLISA will be more reliable. Indeed, MGD Yang-Mills-Dirac stars stellar configurations take into account a finite brane tension, whose current values  comply with the CMB anisotropy and also to well established  cosmological models. For example, the same procedure implemented in the MGD context in Refs. \cite{Fernandes-Silva:2018abr,daRocha:2017cxu}, that predicts the highest frequencies of gravitational wave (GW) radiation emitted by MGD star mergers, and their GW windows to be experimentally detectable in LIGO and eLISA, can be studied for MGD Yang-Mills-Dirac stars.
A really important result is the numerical solution in Fig.~\ref{fi5}, for three different values of the brane tension. The maximal masses of the MGD Yang-Mills-Dirac stellar configurations, as a function of the energy, are smaller, for smaller values of the brane tension. In fact, for any finite value of the brane tension, the MGD Yang-Mills-Dirac stellar configuration maximal mass is smaller than the GR limit, studied in Ref. \cite{Dzhunushaliev:2019uft}. This difference can be observationally explored in current experiments on GW, determining specific physical signatures of MGD Yang-Mills-Dirac stars. 

Beyond the scope of this work, as a relevant perspective, one can add mass to the Yang-Mills gauge potential, to generate MGD Proca stars. Their GR limit were studied in Refs. \cite{Dzhunushaliev:2019uft,Minamitsuji:2017pdr}, with  static, regular, asymptotically flat metric \cite{Brito:2015pxa}.
Current proposals assert that massive spin-1 gauge field can compose dark matter. Hence, MGD Proca stars can play an important role on modelling realistic stellar configurations that are in full compliance with current cosmological models.
Besides, although the MGD Yang-Mills-Dirac stellar configurations have been here scrutinized 
from a first quantized setup, second quantized quantum effects are worthy for investigation.
Also, spin-1/2 fermions were used in this work, but higher spin fermions may be also 
useful to couple to Yang-Mills field and gravity. Besides the spin-3/2 Rarita-Schwinger field already proposed in Ref. 
\cite{Herdeiro:2019mbz}, other fermionic fields, including flagpole, mass dimension one, spinor fields \cite{daRocha:2005ti}, might serve as useful ans\"atze, alternative to the ones in Eq. (\ref{spn1}, \ref{spinor}).

\subsection*{Acknowledgements}

RdR~is grateful to FAPESP (Grant No.  2017/18897-8) and to the National Council for Scientific and Technological Development  -- CNPq (Grants No. 303390/2019-0, No. 406134/2018-9 and No. 303293/2015-2), for partial financial support.

\bibliography{bib_DSS}

\end{document}